\documentclass[a4paper,11pt]{article}
\usepackage{pos}
\usepackage{physics} 
\usepackage{aas_macros} 

\newcommand\rmd{{\rm d}}
\newcommand\aobs{{A_{\rm obs}}}
\newcommand\eobs{{E_{\rm obs}}}

\newcommand\simlt{\lower.5ex\hbox{$\; \buildrel < \over \sim \;$}}
\newcommand\simgt{\lower.5ex\hbox{$\; \buildrel > \over \sim \;$}}

\title{Extreme Energy Cosmic Rays "Treasure Maps": a new methodology to unveil the nature of cosmic accelerators}
 \ShortTitle{Treasure Maps}

\author*[a,b]{Noémie Globus}
\author[c,d]{Anatoli Fedynitch}
\author[e,f]{Roger Blandford}

\affiliation[a]{Department of Astronomy and Astrophysics, University of California, Santa Cruz,\\
  1156 High Street, Santa Cruz, CA 95064, USA}

\affiliation[b]{Astrophysical Big Bang Laboratory, RIKEN, Wako, Saitama, Japan}

\affiliation[c]{Institute of Physics, Academia Sinica, Taipei City, 11529, Taiwan}

\affiliation[d]{Institute for Cosmic Ray Research, the University of Tokyo, 5-1-5 Kashiwa-no-ha, Kashiwa, Chiba 277-8582, Japan}

\affiliation[e]{Kavli Institute for Particle Astrophysics and Cosmology at Stanford University,\\
452 Lomita Mall, Stanford, CA 94305, USA}

\affiliation[f]{SLAC National Accelerator Laboratory,
2575 Sand Hill Road, Menlo Park, CA 94025, USA}

\emailAdd{noglobus@ucsc.edu}

\abstract{
Extreme Energy Cosmic Rays, EECRs -- cosmic rays with energies beyond the GZK cutoff (i.e. greater than 100 EeV) are scarce. Only a few of such events have been detected by air shower experiments and the nature of the primary particles are still unknown. Individual EECRs sources become more prominent, relative to the background, as the horizon diminishes.  We show that   an event-by-event, composition-dependent observatory would allow us to limit the character of the sources and learn about the intervening magnetic fields, as the deflections in the intervening Galactic and extragalactic magnetic fields depend on the nature of the particle. A major goal here is to provide a methodology to distinguish between steady and transient sources.}

\ConferenceLogo{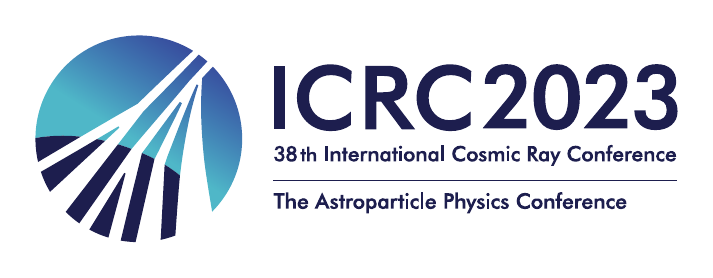}

\FullConference{%
38th International Cosmic Ray Conference (ICRC2023)\\
  26 July - 3 August, 2023\\
  Nagoya, Japan}

\begin{document}
\maketitle

\section{Introduction}

The mystery of the origin of cosmic rays remains surely one of the biggest challenges in high energy astrophysics. These "messengers" of high energy sources  are rapidly isotropized by magnetic waves when the gyroradius is comparable to the coherence length of the turbulent magnetic fields~\citep{2008A&A...479...97G}. Because the intergalactic magnetic turbulence has larger scales of the order of a megaparsec (typically the size of the largest structures in the universe) there was a hope that at the highest energies, the gyroradius would be large enough so that the cosmic rays would travel in straight lines, with their arrival directions pointing to their sources. However, the large scale magnetic fields from our own Galaxy also deflect cosmic rays from their initial directions, with angle of deflections that are quite large even at a rigidity  (momentum per unit charge) of $10^{19}$~V  which corresponds to iron nuclei at 260 EeV (1~EeV$\equiv1\,10^{18}$~eV) \citep{2019JCAP...05..004F}. 

Both Pierre Auger Observatory and  Telescope Array reported a handful of high energy cosmic rays events above 150 EeV (the most extreme energy cosmic rays, EECRs, and the ones we focus on this paper). There are some important benefits of using EECRs, even if such events are rare, to constraint the origin of the sources:
\begin{enumerate}
    \item[{\small $\bullet$}]Because of the limited horizon at the highest energies, there are only a few number of source candidates (see Fig.~\ref{fig:just_sources}). The reason for this is the limited mean free path at the highest energies due to the energy and mass losses by interaction with the ubiquitous Cosmic Microwave Background  as well as the Extragalactic Background Light \citep[][for a review]{Allard2012}. Therefore, the cosmic-ray "horizon" (defined  as the distance from which 95\% of the cosmic rays of a given energy and composition have been lost) is limited to  our local supercluster ($\sim$40 Mpc) for proton and iron nuclei at 150~EeV, which corresponds to our local supercluster.
\item[{\small $\bullet$}]At the highest rigidities, the effect of the Galactic Magnetic Field (hereafter GMF) is better characterized because it is mostly the large-scale halo magnetic field (deduced from Faraday rotation measurements) that affects the cosmic ray transport. Assuming a  GMF  model, it is possible to calculate the deflections and time delays in the Galaxy. The particle trajectory depends on its rigidity. If the nature of the particle is unknown, it is difficult to know the history of its propagation in the Galaxy. This is why it is of prime importance to  have better  constraints on the mass of the cosmic-ray events with future experiments.
\end{enumerate}

We propose here a new methodology, using our current knowledge of the GMF, the attenuation factor due to the GZK effect and the source distribution in the local universe to  build "treasure maps" of the  most promising directions to detect doublets of EECR events in the future \citep{GFB2023}.

\section{GZK horizons and source candidates} \label{sec:gzk-horizon}


We model the source spectrum of the cosmic rays  (denoted by the $s$ index) as a power-law spectrum with spectral index $\gamma$ and a cutoff energy $E_{s, \text{max}}$:
\begin{equation}
\label{eq:source_spectrum}
    \frac{\rmd N_s}{\rmd E_s}(E_s, E_{s, \text{min}}, E_{s, \text{max}}) \propto \left (\frac{E_s}{E_{s, \text{min}}} \right )^{-\gamma} e^{-\frac{E_s}{E_{s, \text{max}}}}, E_s>E_{s,\rm{min}}.
\end{equation}

 The source spectrum from Eq.~\ref{eq:source_spectrum} is repeatedly injected at distances $d_s$ for various choices of $E_{s, \text{min}}$, $E_{s, \text{max}}$, $\gamma$, and the several nucleus masses from $A_s=1$ (protons) up to $A_s=56$ (iron). 
 
 We introduce the parameters $\eobs$ and $\aobs$, which are the lower thresholds for the energy and mass number of cosmic rays observed at Earth. The extragalactic propagation of EECR between the source location and the Earth is modeled using the code {\sc PriNCe} \cite{Heinze:2019jou}. We define the spectrum at the distance of the Earth as
\begin{equation}
    \frac{\rmd T_{A_i}}{\rmd E}( d_s, \gamma, E_{s, \text{min}}, E_{s, \text{max}})
\end{equation}
separately for each nuclear mass $A_i \leq A_s$. The lower masses are populated due to the disintegration of nuclei in photo-nuclear and photo-hadronic interactions.

 
 We work towards the definition of the GZK horizon using the \textit{loss of number}, in analogy to an attenuation coefficient:
\begin{equation}
\begin{split}
\label{eq:agzk}
    a_\text{GZK}(&A_s, d_s, E_{s, \text{max}}, \gamma~|~\aobs, \eobs) =\\ &=\frac{\sum_{A_i \geq \aobs}\int_\eobs^\infty \rmd E \frac{\rmd T_{A_i}}{\rmd E}( d_s, \gamma, E_{s, \text{max}})}{\int_\eobs^\infty \rmd E_s \frac{\rmd N_s}{\rmd E_s}(E_s, E_{s, \text{max}})},
\end{split}
\end{equation}
setting $E_{s, \text{min}}$ to $\eobs\geq150$~EeV. The resulting $a_\text{GZK}$ are then calculated  for several choices of $\aobs$, $\gamma$ and $A_s$. 
Any observed high energy cosmic-ray proton can be the remnant of a heavier nuclei starting its journey at larger distances than the strict horizon for protons. 
The EECR horizon is limited to  our local supercluster ($\sim$40 Mpc) for EECRs starting as proton and iron nuclei at 150~EeV. For intermediate mass cosmic rays such as CNO, the horizon is not larger than $\sim2$~Mpc, i.e. limited to the size of the local group. The observational capability for the mass composition above 100 EeV thus carries critical information about the EECR source distance. 

\begin{figure*}
\centering
\includegraphics[width=0.49\textwidth]{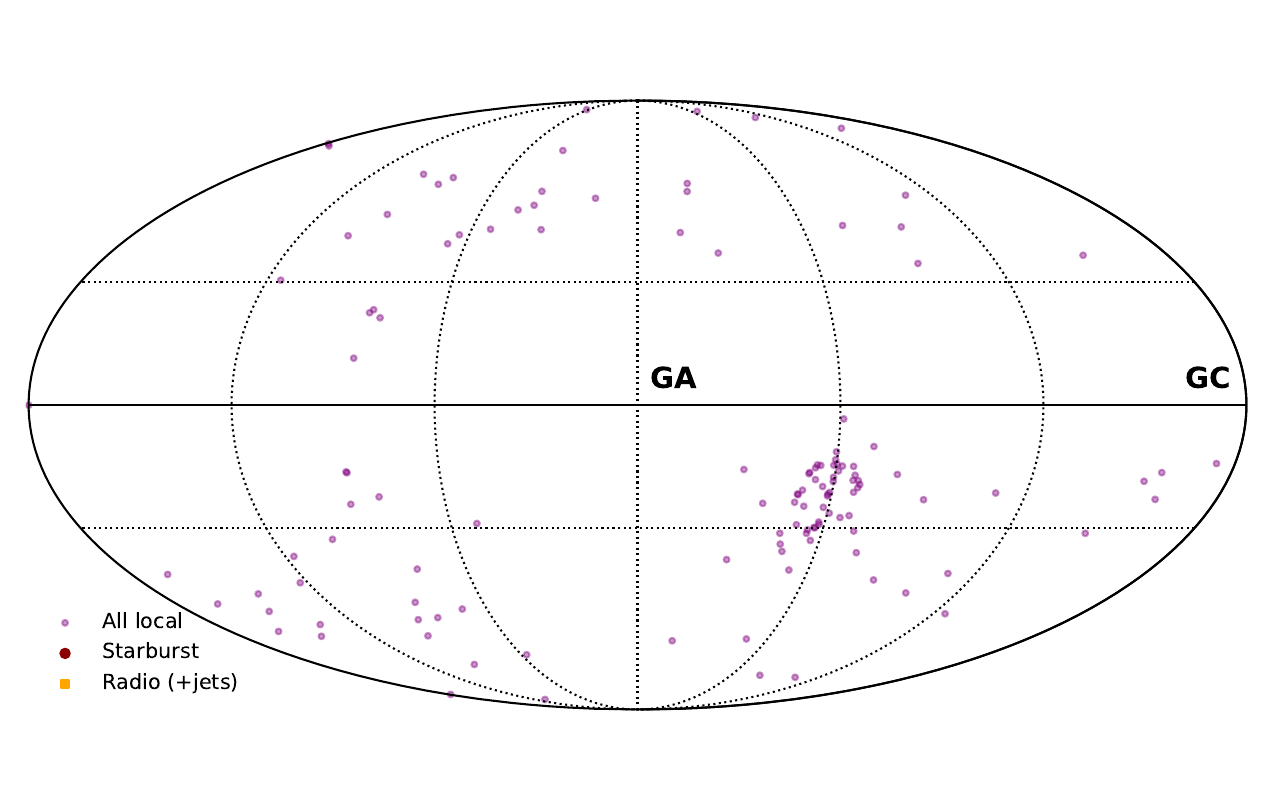}
\includegraphics[width=0.49\textwidth]{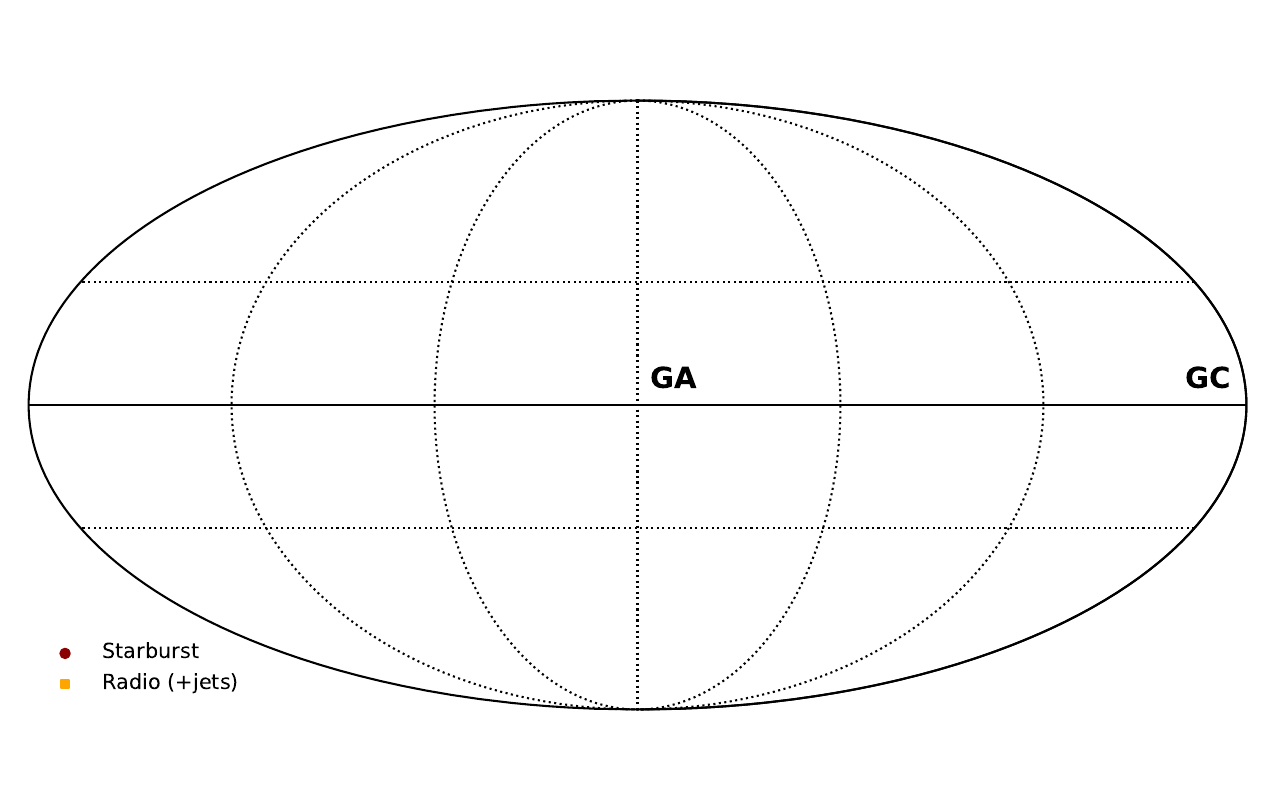}
\includegraphics[width=0.49\textwidth]{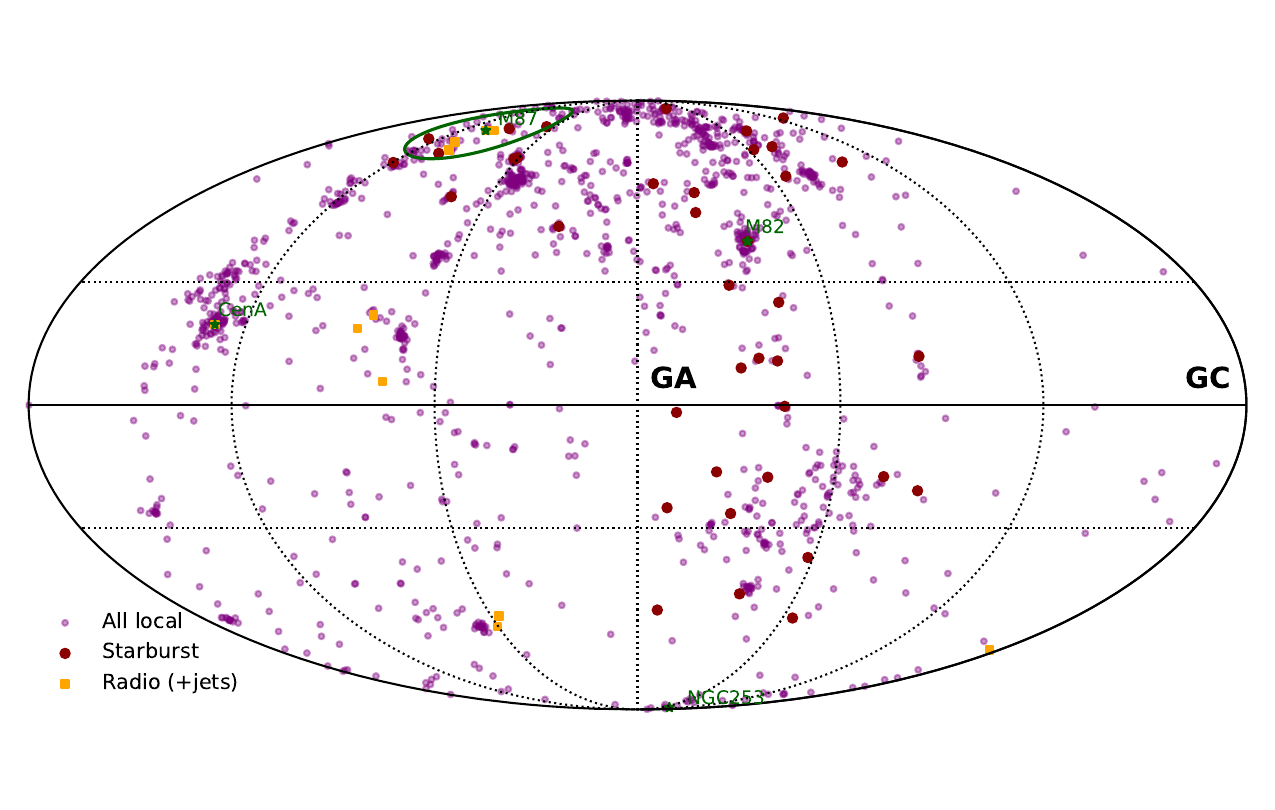}
\includegraphics[width=0.49\textwidth]{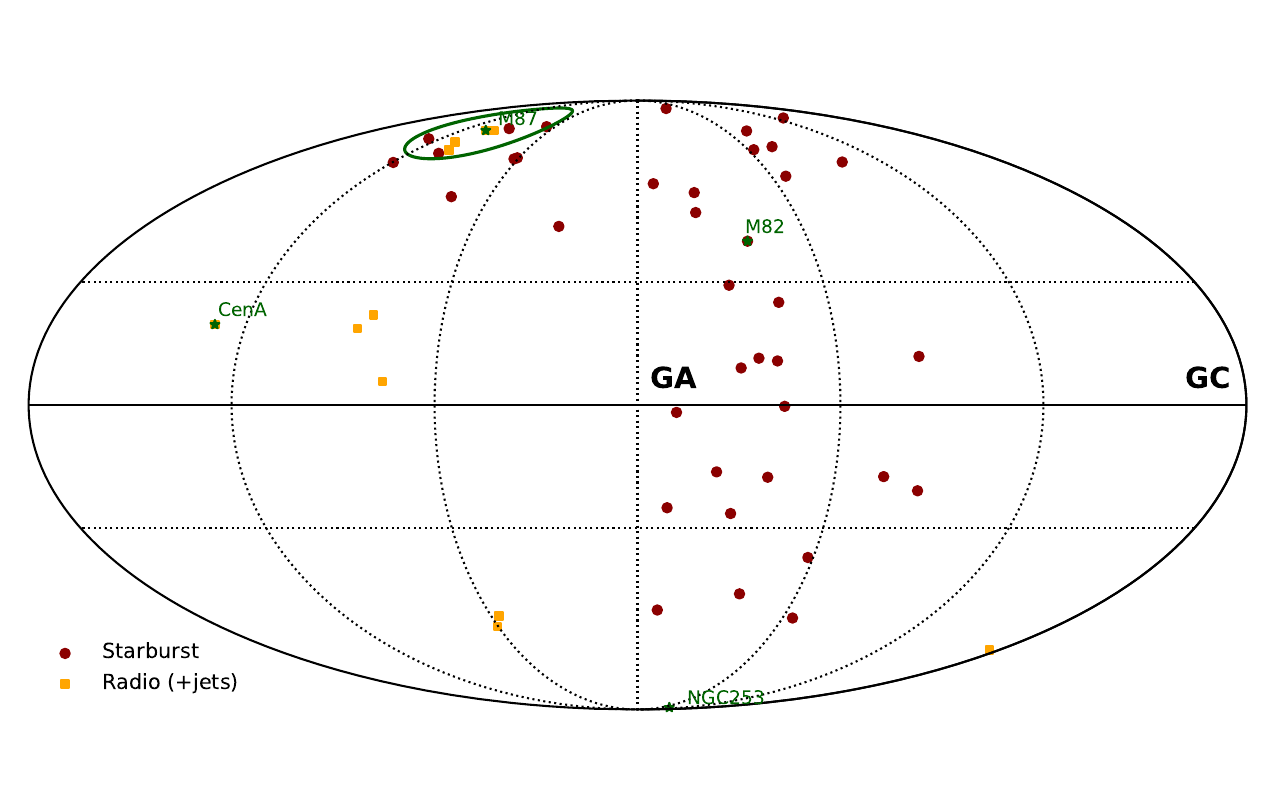}
\caption{Source candidates within 2 Mpc (top) and 40 Mpc (bottom). On the left panel, the maps contain all galaxies, Starbursts and Jetted AGNs;  while on the right panel, only  Starbursts and Jetted AGNs are shown.  Some popular candidate sources M82, Centaurus A, NGC253, and M87 (including a $10^\circ$ circle for the expected accretion shock) are shown as green stars to aid with orientation.  }
\label{fig:just_sources}
\end{figure*}

We use the Updated Nearby Galaxy Catalog  \citep{Karachentsev:2013cva}. As of 2022, the catalog includes 1,420 galaxies with an sufficiently precise distance estimate. With the exception of KK198, which is located at a distance of 49 Mpc, all other galaxies are within a distance of 30 Mpc. We note that 85\% of the entries of this catalog are dwarf galaxies, which follow a similar distance distribution as regular galaxies. For jetted AGNs we use the a subset of the catalog \citep{vanVelzen2012} and for Starburst Galaxies (SBG) the catalog  \citep{2019JCAP...10..073L}. The sources candidates from these different catalogs are shown in Fig.~\ref{fig:just_sources}.   One can see that for EECRs with $\eobs\geq150$~EeV, $\aobs\geq 12$ there are no Starburst or Jetted AGN host candidates, while for for EECRs with $\eobs\geq150$~EeV, $\aobs\geq 1$, there are Starburst, Jetted AGNs and Cluster Accretion Shocks within the EECR horizon.

\section{Magnetic Fields}

\subsection{Magnetic Lensing in the Galaxy}\label{lensing}
The GMF is responsible for the point source magnification $M$. The large scale field can both increase and decrease the cosmic-ray flux through magnetic focusing/lensing \citep{harari2000}.  The turbulent magnetic field splinters the particle trajectories into ever-diminishing regions spreading over an increasing area and leads to pitch angle scattering. It is the second effect that is most important for determining the magnification so long as the deflections are small; when the deflections are large, the cosmic rays from a given sources will be mostly affected by the large-scale, coherent field. Thanks to the recent progress made in the modeling of the  GMF, we have been  able to calculate the magnification factors $M$ {\color{black}assuming a specific GMF model. 

\subsection{Temporal dispersion in Galactic and extragalactic magnetic fields}\label{dispersion}

The temporal dispersion $\tau_d$ for two cosmic rays with the same rigidity, has two contributions, one  from the GMF, $\tau_{d,\text{GMF}}$, which we calculate  assuming a GMF model, and one from  the extragalactic magnetic field (EGMF), $\tau_{d,\text{EGMF}}$. Assuming a purely turbulent extragalactic field with a Kolmogorov spectrum with a rms intensity $B_{\rm nG}$ and a coherence length $\lambda_{\rm Mpc}$, the average extragalactic dispersion is  
    $\tau_{d, {\rm EGMF}}\sim 5\, {\rm yr}\, (E_{200}/Z)^{-2} B_{\rm nG}^2 {d_s}_{\rm Mpc}^2 \lambda_{\rm Mpc}$ \citep{Lemoine1997}.

Based on the available GMF models, we can provide an estimate of the temporal dispersion in the Galaxy as a function of the direction of the source in the sky and the rigidity. We  use the Jansson and Farrar GMF model \citep{JF12}, with and without the correction introduced with Planck observations \citep{Planck2016}   and the Terral and Ferri\`ere GMF model \citep{TF17} to calculate the temporal dispersion of the EECR signal due to the GMF  as a function of the source direction and the rigidity of the particle. The propagation of cosmic rays through the GMF is modeled using the  Monte Carlo code CRPropa 3.2 \citep{AlvesBatista:2022vem}. We refer the reader to \citep{GFB2023} for the details of the calculation. Multiplets of EECR events are more likely to be detected in ``magnetic windows'' of the sky where the temporal dispersion, $\tau_d$, is small enough to detect at least one doublet within the typical observation time of an EECR observatory, typically decades. This window has to back-project to   source candidates, which is not guaranteed when the sources are scarce due to the limited horizon.

\section{Detecting EECR doublets}

\subsection{General considerations}
Given the apparent advantages of EECRs, we are interested in characterizing how many source candidates from our local volume (or local supercluster) can be expected to be accessible. To answer this question one must consider several arguments:
\begin{enumerate}
    \item[{\small $\bullet$}] The source distances are limited by the attenuation functions $a_\text{GZK}$ due to the GZK effect, see Sec.~\ref{sec:gzk-horizon}. Thus, the number of candidate sources  depends on the choice of $\aobs$, $\eobs$ and $A_s$, and to some extent on the assumed source parameters, like $E_{\rm max}$ or $\gamma$.
    \item[{\small $\bullet$}] Transients sources, such as gamma-ray bursts or tidal disruption events, are, presumably, correlated with the distribution of galaxies. 
    \item[{\small $\bullet$}] Due to the deflections within the GMF, parts of the sky are inaccessible to observations since there are no valid trajectories that can connect the Earth to a source. On the other hand, magnetic lensing can magnify certain directions, as explained in Sec.~\ref{lensing}.
    \item[{\small $\bullet$}] Due to the turbulence of the GMF, cosmic rays arriving from the same direction have slightly different travel times leading to a temporal dispersion $\tau_{d,\text{GMF}}$. 
    It has to be combined with the dispersion due to the EGMF, $\tau_{d,\text{EGMF}}$. These are explained in Sec.~\ref{dispersion}.
    \item[{\small $\bullet$}] One needs to consider the location and zenith angle coverage in case of a ground based detector. In particular, this is important for heavier compositions where the accessible regions of the sky don't coincide with the exposure function at Earth, see Fig.~\ref{fig:TM_150} for illustration.
\end{enumerate}

\subsection{Number of host galaxy candidates for detecting EECR doublets}

For a transient source at 50 $d_{50}$ Mpc to create a doublet of 2 EECR events at $200 E_{200}$ EeV, the minimum isotropic equivalent energy (denoted as $U_{\rm iso,2}$) would need to be 
\begin{equation}
U_{\rm iso,2} \sim 4.38 \cdot 10^{52}{\rm erg}\, (\tau_d/10^3 {\rm yr})d_{50}^{2}{\,} E_{200}^{\,} {({\cal E}/{\cal E}_{\rm PAO})}^{-1} a_{\rm GZK}^{-1} M^{-1} n_{\rm yr}^{-1} \,,
\label{eq:u_iso}
\end{equation}
where $n_{\rm yr}$ is the number of years of observation. For TA, the annual exposure is ${\cal E}_{\rm TA} \sim 900\,\, \rm{km^2\, sr\, yr}$, so the isotropic equivalent energy  needed would have to be  $\sim$six times higher than for PAO. The number of host galaxies within the field of view that can harbor transient EECR sources able to provide a doublet is performed using a weighted sum, where for each source in the catalog the weight is defined as
\begin{equation}
  \hat{w}_s = \max{\left(1, \frac{U_{\rm iso}}{U_{\rm iso,2}} \right)}\,,
\label{eq:w_count}
\end{equation}
where $U_{\rm iso,2}$ is given by Eq.~\ref{eq:u_iso} (the energy necessary to provide a EECR doublet after taking GZK and magnification into account) and $U_{\rm iso}$ is the true source energy (which can be larger or smaller).  
The weights are capped at 1 for the doublet case,  so once a catalog source reaches this condition it is fully counted, whereas a source doesn't contribute if it is not within the field of view ($M=0$) or beyond the GZK horizon ($d_s \geq d_{95\%}$). Transients within host galaxies with $\hat{w}_s$ between 0 and 1 can produce observable doublets with reduced probability.  

\subsection{''Treasure Maps'' of the most promising directions for  detecting  EECR doublets}\label{sec:maps}

The sky maps, shown in Figs.~\ref{fig:TM_150}  aim to visualize and compress the results of the simulations. The maps are in Galactic coordinates and centered on the Galactic anti-center (GA), since for EECRs heavier than protons the directions behind the Galactic center (GC) are unreachable due to the strong deflections, and hence the maps are often blank there. Color coding is used for the time dispersion $\tau_{d, \text{GMF}}$. Transparency is assigned to the magnification map, where $M=0$ corresponding to fully transparent and $M\geq 1$ to solid color. {\color{black}The host galaxy candidates are shown as circles, their} color corresponds to the total $\tau_{d}$, and {\color{black}their} transparency channel is set to $\min(a_\text{GZK}, M)$. Sources farther than $d_{95\%}$ are not shown for clarity. We refer the reader to Fig.~\ref{fig:just_sources} where some popular candidates sources are displayed for orientation.

These treasure sky maps  demonstrate that to estimate the visibility of specific host galaxies within the field of view of an observatory requires anisotropic, four-dimensional modeling (in direction, distance, and temporal dispersion). The GMF plays a crucial role even at these extreme energies. Thus, neglecting the GMF's impact or approximating deflections by isotropic smearing kernels is inappropriate for source searches under the assumption of light or heavy nuclear composition. We have compiled the figures for all  the tested models in a public data release \citep{EECR_treasure_maps_repo}.

\section{Summary}

The minimum isotropic energy $U_{\rm iso}$ in EECRs above 150 EeV to get a doublet from a single transient source at a distance $d_s$, in the direction $(l,b)$, depends on the total temporal dispersion $\tau_d$ (in the EGMF and the GMF), on the GZK attenuation $a_{\rm GZK}(d_s)$, on the magnification $M(l,b)$, and the field of view of the experiment, all displayed in the treasure maps. We then count the number of sources satisfying this condition. For the most optimistic case where the EGMF is negligible, we get 100 galaxies satisfying this condition for $U_{\rm iso}=10^{47}$~erg, and 1,000 for $U_{\rm iso}>10^{49}$~erg. So, even if we have 1000 candidate galaxies, we have to take into account the rate of the transients to estimate the probability to observe a doublet during 10 year observation period, typical for an UHECR observatory. The minimum rate is then $10^{-4}$yr$^{-1}$galaxy$^{-1}$. We can compare this rate to the rates of known transient sources, and we refer the reader to our paper for this discussion \citep{GFB2023}.

We also showed that an analysis of the multiplet arrival times would allow us to distinguish between transient and continuous sources. Essentially, time arrival of events with a similar rigidity can be drawn independently from the asymmetrical distribution of time delays and compared with the data. The discovery of more EECRs and multiplets of events with a composition-sensitive detector could rule out many currently viable source models.

\section*{Acknowledgements}
We thank Alan Watson for useful comments.
N.G.’s research is supported by the Simons Foundation, the Chancellor Fellowship at UCSC and the Vera Rubin Presidential Chair. A.F.\ acknowledges the hospitality within the group of Hiroyuki Sagawa at the ICRR, where he completed initial parts of this work as a JSPS International Research Fellow (JSPS KAKENHI Grant Number 19F19750). We acknowledge the computational resources and support provided by the Academia Sinica Grid-Computing Center (ASGC), which is supported by Academia Sinica.

 \begin{figure*}
\centering
\includegraphics[width=0.45\textwidth]{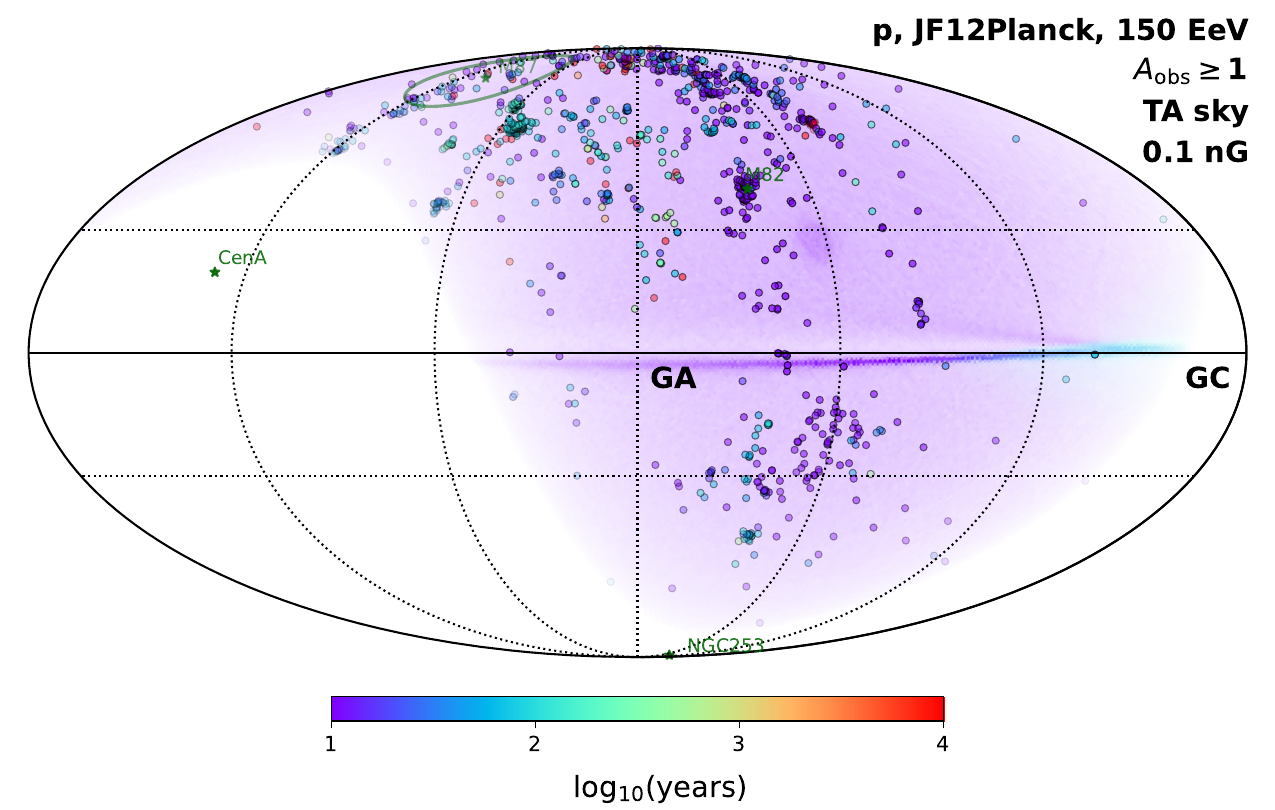}
\includegraphics[width=0.45\textwidth]{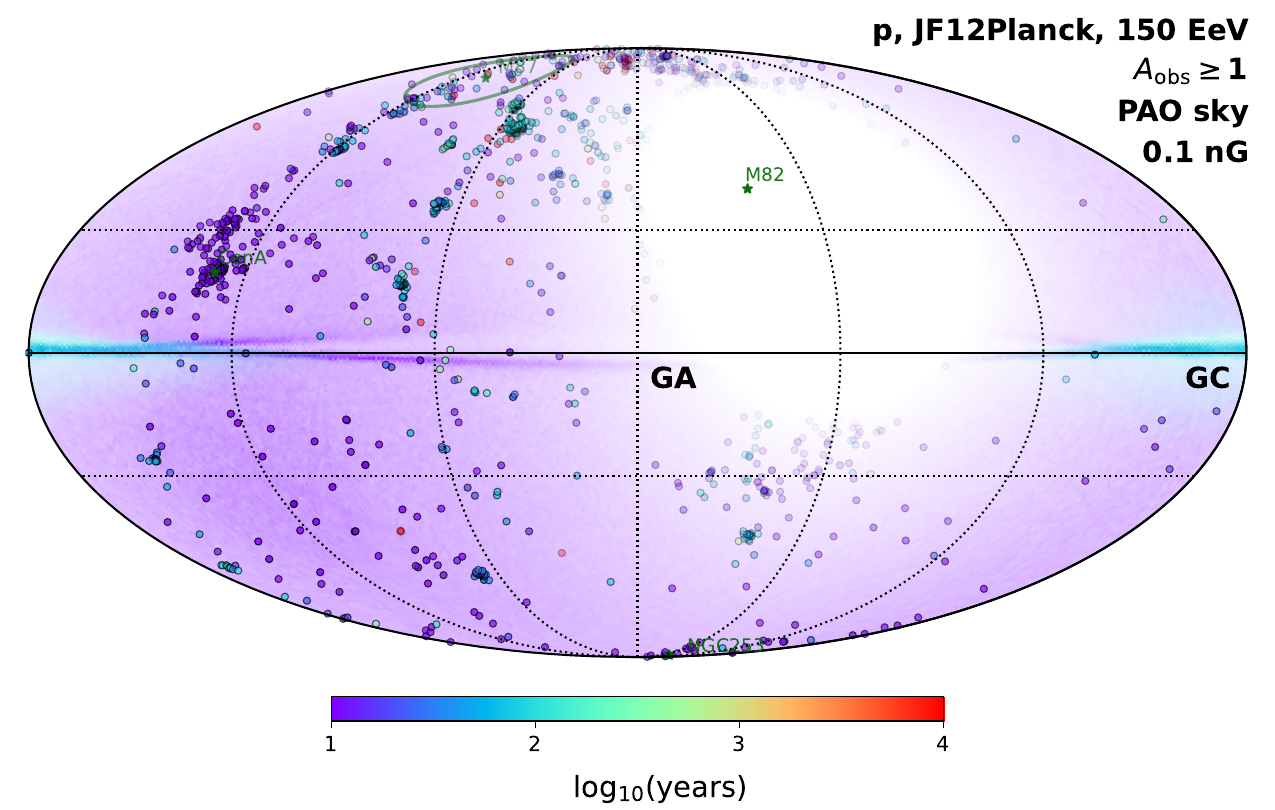}
\includegraphics[width=0.45\textwidth]{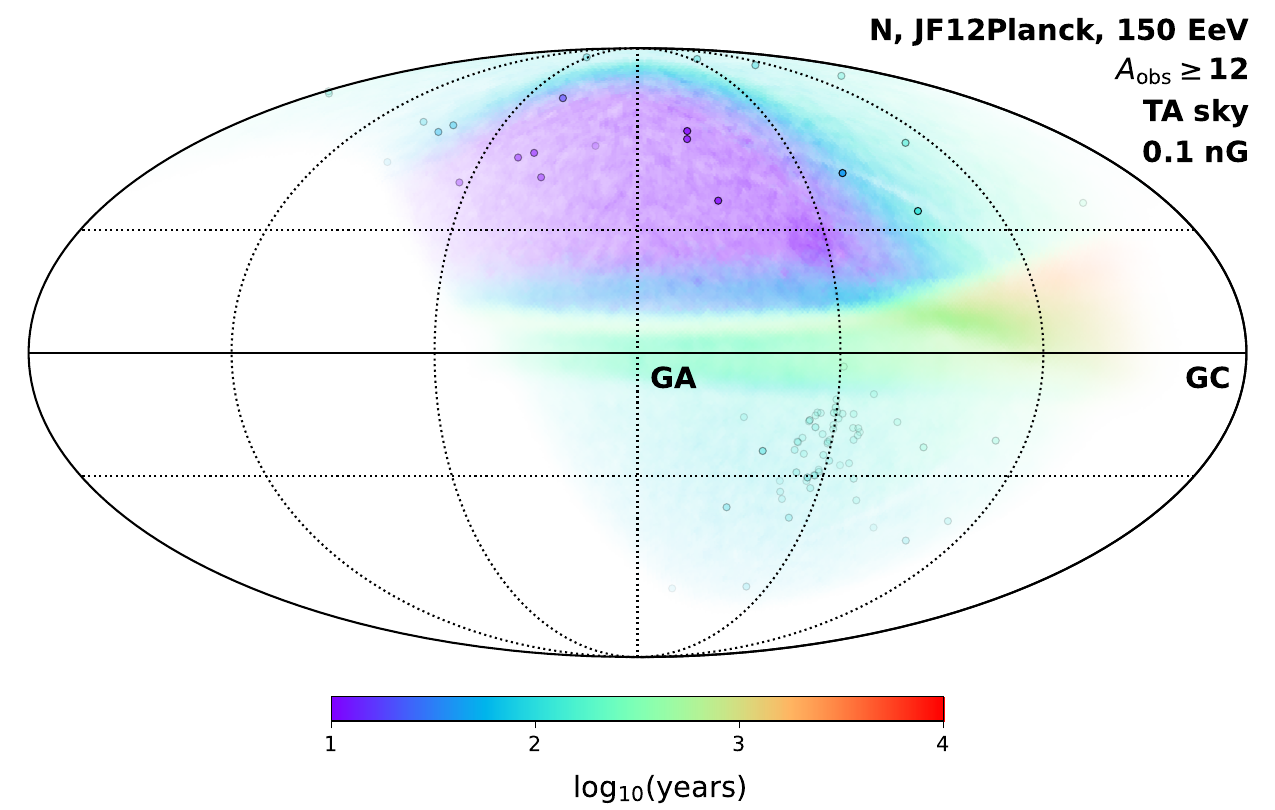}
\includegraphics[width=0.45\textwidth]{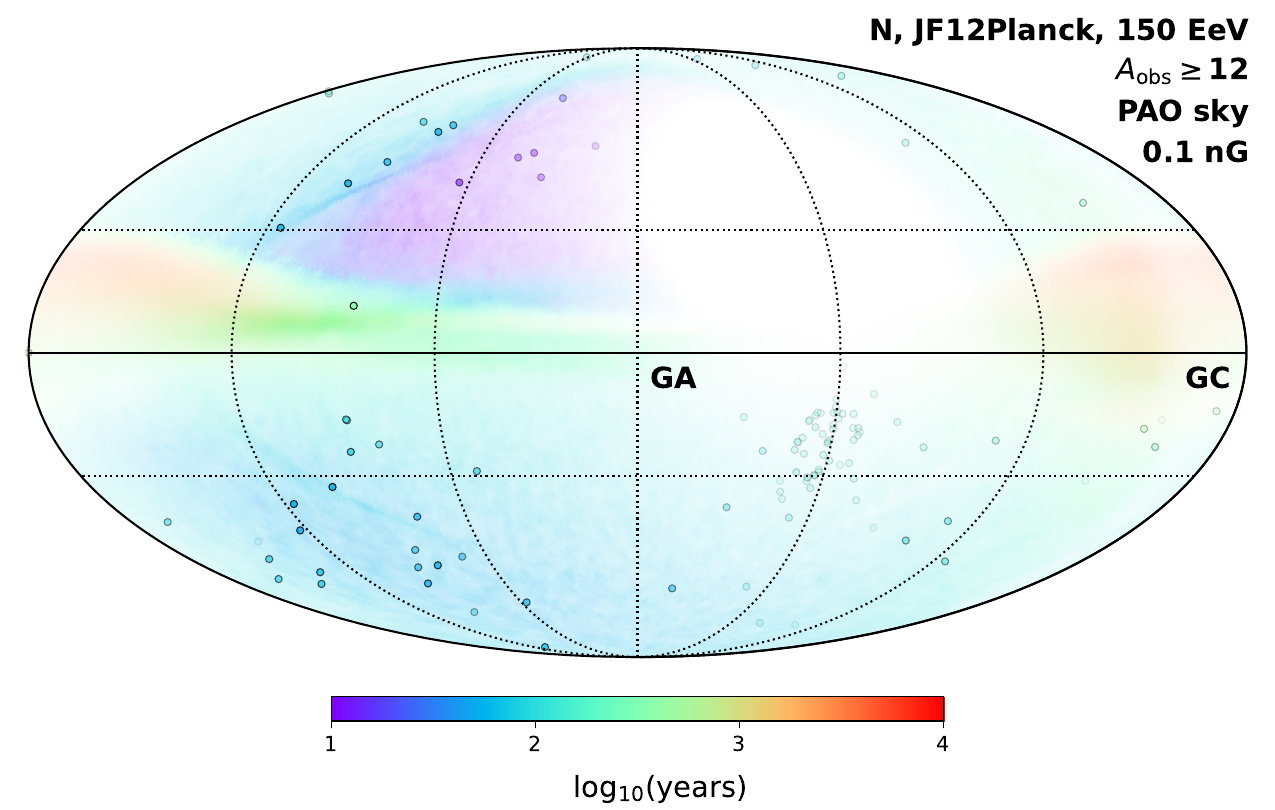}

\includegraphics[width=0.45\textwidth]{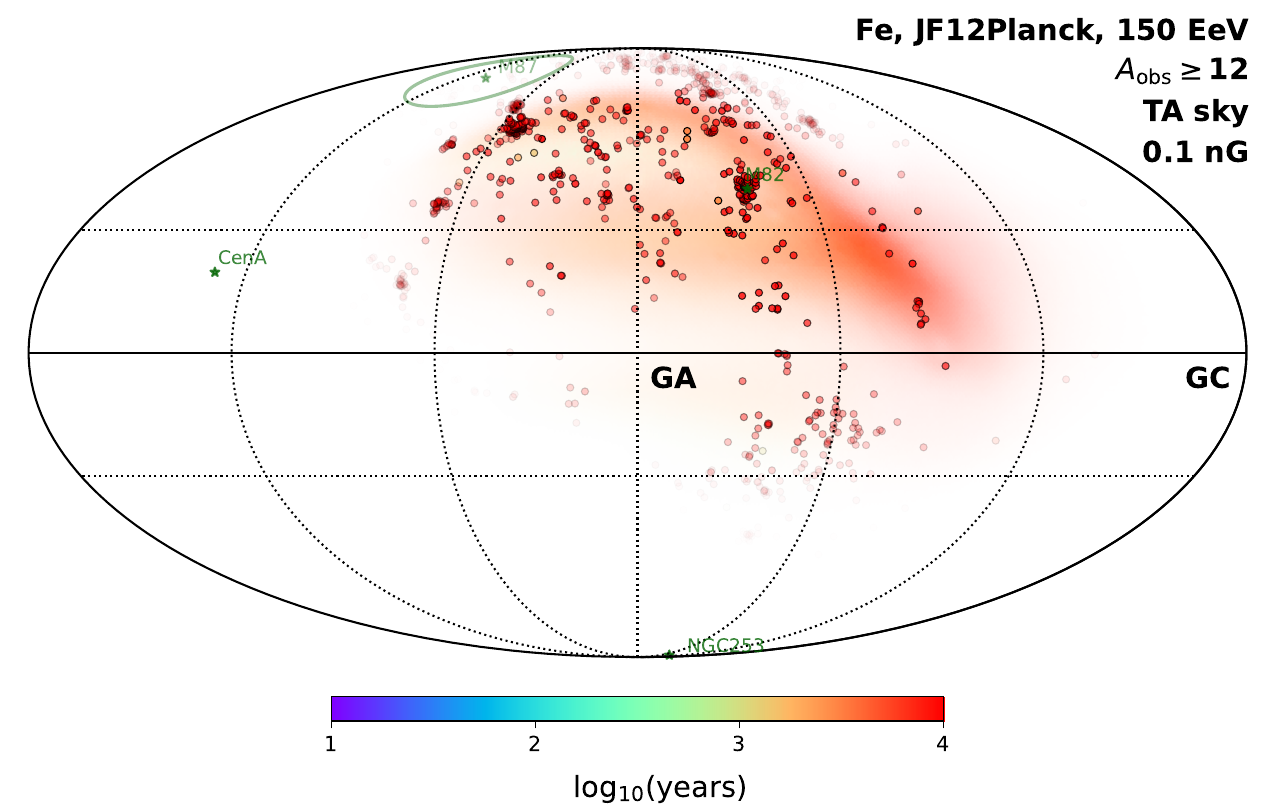}
\includegraphics[width=0.45\textwidth]{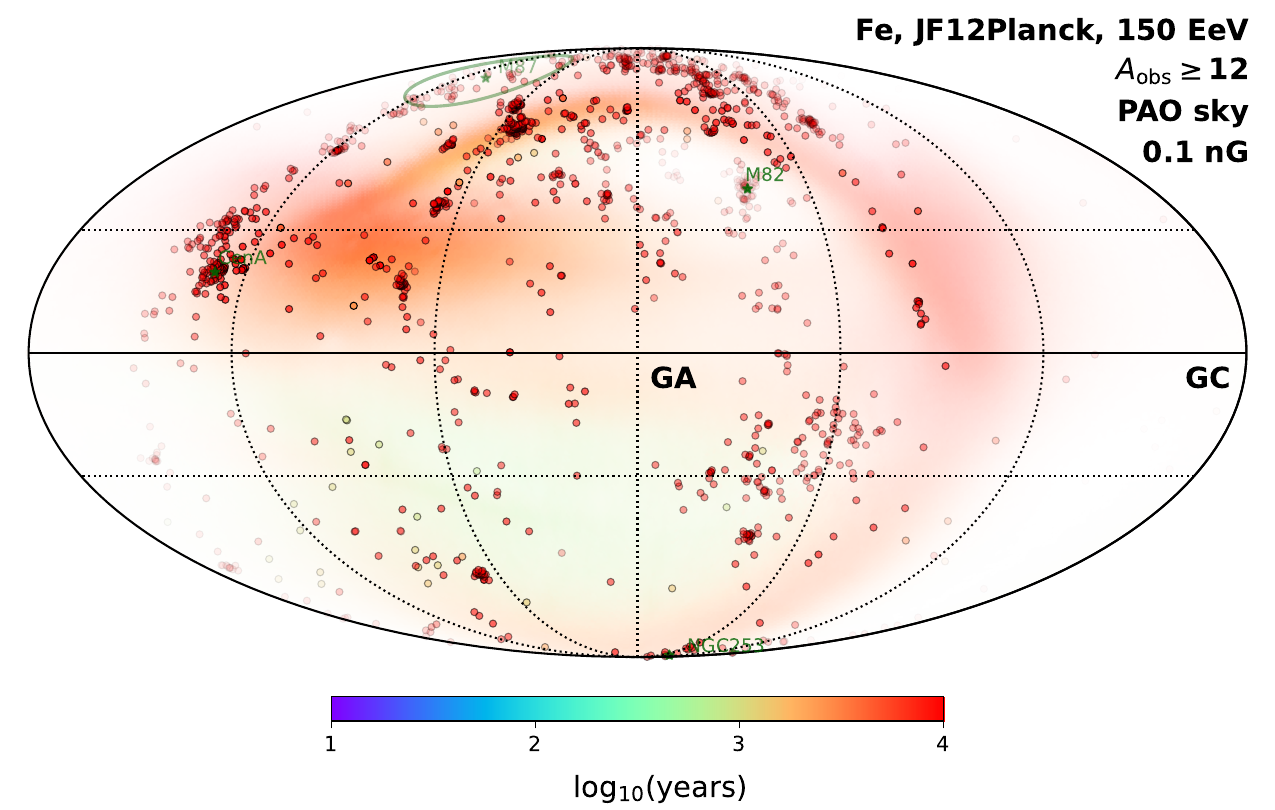}

\caption{EECRs at 150 EeV  ``treasure maps'' for TA (left) and PAO (right), for protons (top),  nitrogen (middle), iron (bottom) for the JF12Planck  GMF model. The color of the gradients is assigned to $\tau_{d, \text{GMF}}$. The gradient's opacity is controlled by the truncated magnification maps $\min(M, 1)$, which include the detector exposure function. The source colors are assigned to the total $\tau_{d}$ (the EGMF strength is {\color{black} 0.1 nG} as indicated). The source marker opacity is set to $\min(a_\text{GZK}, M)$, i.e.~within the detector's exposure the markers fade mostly due to $a_\text{GZK}$ and due to $M$ outside of it. We consider here the case $\aobs\geq 12$ (i.e., at detection), to demonstrate the impact of on the observed volume in case of a composition-sensitive detector that can provide a sub-sample of nucleus-like events.}
\label{fig:TM_150}
\end{figure*}

\bibliographystyle{JHEP}
\bibliography{refs}

\end{document}